\documentclass[showpacs,twocolumn,amsmath,superscriptaddress,pre]{revtex4}
\usepackage[dvips]{graphicx}
\usepackage{epsfig}
\usepackage{color}
\usepackage{subfigure}
\usepackage{soul}

\begin{document} 

\title{
Moments of undersampled distributions:
Application to the size of epidemics
%Finite-size scaling 
%versus dual random variables and
%shadow moments
%in the size distribution of epidemics
%%Scientific comment on 
%%``Tail risk of contagious diseases''
}
\author{\'Alvaro Corral}
\affiliation{%
Centre de Recerca Matem\`atica,
Edifici C, Campus Bellaterra,
E-08193 Barcelona, Spain
}\affiliation{Departament de Matem\`atiques,
Facultat de Ci\`encies,
Universitat Aut\`onoma de Barcelona,
E-08193 Barcelona, Spain
}\affiliation{Complexity Science Hub Vienna,
Josefst\"adter Stra$\beta$e 39,
1080 Vienna,
Austria
}
%\section*{Abstract}
%
%
\begin{abstract} 
The total number of fatalities of an epidemic outbreak is a dramatic but extremely informative quantity.
Knowledge of the statistics of this quantity allows the calculation of the 
mean total number of fatalities conditioned to the fact that the outbreak has surpassed a given number of fatalities, which is very relevant for risk assessment.
However, the fact that the total number of fatalities seems to be characterized by a power-law tailed distribution with exponent (for the complementary cumulative distribution function) smaller than one poses an important theoretical difficulty, due to the non-existence of a mean value for such distributions.
Cirillo and Taleb [Nature Phys. 16, 606 (2020)] propose a transformation from a so-called dual variable, which displays a power-law tail, to the total number of fatalities, 
which becomes bounded by the total world population.
Here, we
(i) show that such a transformation is ad hoc and unphysical;
(ii) propose alternative transformations and distributions (also ad hoc);
(iii) argue that the right framework for this problem is statistical physics,
through finite-size scaling;
and
(iv) demonstrate that the real problem is not the non-existence of the mean value for power-law tailed distributions but the fact that the tail of the different theoretical distributions 
(which is what distinguishes one model from the other) is far from being well sampled with the available number of empirical data.
Our results are also valid for many other hazards displaying (apparent) power-law tails in their size.
\end{abstract} 
\maketitle

\section{Introduction}

%COVID: Avalancha de articulos oportunistas!! 
%Este paper es un critica a ello??

The study of 
epidemics has been of the interest of mathematicians and physicists since long ago
%\cite{Lipsitch,Lloyd_superspreading}
\cite{Bernoulli_smallpox,Dietz_Bernoulli,Pastor_Vespignani,Kleinberg_book,Pastor_rmp,Allen,Miller_pgf,Kucharski_book,Hill_epidemics};
%\cite{pausch2021noise}
however, the recent and (at the time of submitting this paper) 
still ongoing  COVID-19 pandemic has triggered an enormous increase
%in the number of publications studying 
in the research about
mathematical or physical aspects of epidemics.
Although the majority of works are related to modeling epidemic spreading
(e.g., \cite{Arenas_covid,Castro_Ares_PNAS20,Falco_Corral}), 
the statistics of epidemics is also of capital importance for risk assessment.
In a suggestive paper, Cirillo and Taleb \cite{Cirillo_Taleb}
%(see also...???)
deal with the relevant issue of the size of epidemic outbreaks, 
understanding size as total (i.e., final) number of fatalities.
For instance, the size of the infamous Black Death, in the 14th century (A.D.), 
was assumed to be 137,500,000 fatalities \cite{Wikipedia_epidemics}.

From their analysis of the data of 72 epidemics gathered mainly in Ref. \cite{Wikipedia_epidemics},
the authors of Ref. \cite{Cirillo_Taleb} claim that 
the distribution of fatalities in historical epidemics seems to be ``extremely fat-tailed'',
which means that the 
complementary cumulative distribution function
(ccdf, also called survival function \cite{footnote1_comment2_CT})
of the total number of fatalities
decays as a (decaying) power law (pl) asymptotically
(i.e., for very large total number of fatalities).
In a formula, 
$$
S_{\text{pl}}(x) \sim \frac 1 {x^{\alpha}},
$$
where 
$x$ is the (total) number of fatalities in each epidemic,
$S(x)$ is its ccdf, 
$\alpha>0$ is the exponent of the ccdf,
which is related to the tail index $\xi$ (measuring fat-tailedeness)
by $\xi=1/\alpha$, 
and the symbol ``$\sim$'' denotes 
%asymptotically proportional.
asymptotical proportionality,
which means that the power law only holds at the ``tail'' of the distribution.
%To avoid confusion, let us mention that
%in survival analysis $S(x)$ accounts for the lifetime $x$ of an individual,
%but this is not the case here;
%nevertheless, the idea is essentially the same, 
%as one counts the ``lifetime'' of an epidemic not using time but number of fatalities instead.
%
%pendiente!! MENCIONAR LA SLOWLY VAR FUNC!!
%

Taking the derivative of $S(x)$, 
one obtains the equivalent description in terms of the probability density $f(x)$, 
which in the case of a power-law tail
fulfills
$$f_{\text{pl}}(x) \sim \frac 1 {x^{1+\alpha}}.$$
%with $f(x)$ the probability density.
In a more formal setup, to avoid the use of the symbol ``$\sim$''
one may introduce a slowly varying function that multiplies the power-law part of $S(x)$,
yielding what is called a regularly varying function
\cite{Voitalov_krioukov}.

The ``extreme fat-tailedness'' of the epidemic fatalities is supported 
in Ref. \cite{Cirillo_Taleb}
by the claim
$\xi=1.62$,
and so, $\alpha=0.62$
%\cite{Cirillo_Taleb}.
(the coincidence of these figures with the golden mean
should be considered serendipitous).
%For a criticism of the power-law-tail perspective of Ref. \cite{Cirillo_Taleb} see 
Although it is not relevant for our purposes, 
for the sake of completeness it is worth to mention that 
the claim of the existence of a power-law tail in the distribution of epidemic sizes \cite{Cirillo_Taleb}
has been criticized in Ref. \cite{Corral_epidemics_pre}, 
where other alternative distributions were proposed.

%PENDIENTE!! 0: LOS 2 PARRAFOS DE ABAJO ESTAN MUY DESORDENADOS

The authors of Ref. \cite{Cirillo_Taleb} correctly mention 
that power-law-tailed distributions are characterized by the lack of moments;
from an idealized ``physical'' point of view
this means that the moments that do not exist can be considered as taking an infinite value.
In concrete,
the $n-$th order moment diverges if $n\ge \alpha$ and,
in particular, not even the first moment (the mean) is finite if 
$\alpha\le  1$.
%
%%%This non-existence of moments 
%The same authors further state the obvious fact that the lack of moments
%for a variable such as the number of fatalities
%is questionable.
%%at least for the distribution 
%%in the case of the number of fatalities in epidemics.
%%
%
%Of course, the random variable $x$ is bounded by the total world population,
%and therefore, the moments cannot be infinite.
%(infinite moments correspond to what mathematicians label as lack of moments, in this case).
%
%
%\section{Finite-size effects}
%
%%The existence or not of moments is an issue that has been debated.
%%On the one hand, although 
%%the (truncated) log-normal distribution is heavy tailed, it is ``well behaved''
%%in the sense that all its moments exists.
%%%(and the same for the truncated log-normal, as the truncation is from below).
%On the other hand, power-law tailed distributions and fat-tailed distributions
%lack all moments of order higher than $\alpha$
%(the zeroth-order moment always exists as $\alpha > 0$).
%In particular, for $\alpha < 1$, no moment exists, not even the mean  
%(the moments can be considered infinite,
%except the zero moment, which is alway one).
%%
%
%Notice that
%
From the point of view of probability theory there is no fundamental impediment in the fact that the moments become infinite, although the law of large numbers and the central limit theorem do not hold 
when $\alpha$ takes the lowest values and one has to rely on the generalized central limit theorem instead \cite{Bouchaud_Georges,Mantegna_Stanley}.
From a practical point of view, one can always calculate an empirical mean value,
the key question is that, when $\alpha<1$, 
%all the moments are infinite, including the mean, and 
the empirical value does not keep ``stable''  but grows as the number of data grows; when $1 <\alpha < 2$ the mean is not infinite but its emprirical estimation shows wild fluctuations.
%
%Obvioulsy, from the physical perspective, this infinitude of moments can seem embarrassing.
%as the number of fatalities is bounded by the total world population,
%and therefore, the moments cannot be infinite.
Naturally \cite{Cirillo_Taleb}, the fact that the number of fatalities is bounded by the total world population
makes absurd the use of power-law tailed distributions to calculate moments
(although still these distributions can be very useful to fit finite sets of data).

%PENDIENTE!! 1: MEZCLA DE COSAS: UNA COSA ES QUE LOS MOMENTOS EMPIRICOS ESTEN MAL DEFINIDOS EN UNA POWER LAW, OTRA QUE EL MOMENTO REAL NO SE CORRESPONDE CON EL MOMENTO TEORICO EN UNA PL TRUNCADA O TAPPEREADA SI LA "COLA" ESTA "LEJOS"
%

In order to be able to calculate the moments of a distribution in a case like this,
%To avoid problems with the moments of the distribution,
the authors of Ref. \cite{Cirillo_Taleb} propose a transformation of the variable $x$
[Eq. (\ref{dual}) below]
that still leads to a power law, with the same exponent, for small and intermediate values of $x$ but that leads to a zero probability of $x$ being larger than the world population $h$.
The same approach has been previously applied to other systems which were yielding apparent power-law distributions \cite{Cirillo_Taleb_wars,Cirillo_Taleb_QF}.
%, apart from epidemics.
On the other hand, it is obvious that the problem of the lack of moments in the size of epidemics is not present if one takes the approach of Ref. \cite{Corral_epidemics_pre}, 
where it was shown that there are other distributions that fit the epidemic data better than a power-law tail, 
and for these distributions (e.g., the truncated lognormal) all the moments exist.

We will treat the problem of the lack moments 
%of power-law tailed distributions 
from a theoretical generic point of view, without paying special attention to the fact of which is the best distribution to fit the size of epidemics \cite{Corral_epidemics_pre} or any other concrete system \cite{Cirillo_Taleb_wars,Cirillo_Taleb_QF,Clauset,Corral_Deluca,Corral_Gonzalez}.
Nevertheless, we will consider the case of the size of epidemics as a working example.

A fundamental outcome of our research is that the main problem is not the divergence of moments
when there are power-law tails, 
but the fact that the part of the distribution corresponding to the real tail
(the largest values in practice) are not well sampled with the available data 
and the values of moments cannot be established, 
no matter the model one uses.
In other words, the resulting moments depend on the selected model,
but the available data does not allow to discern between different models at the tail.

This paper is organized as follows:
In the next section we introduce the original approach of Ref. \cite{Cirillo_Taleb}
and present some alternative transformations and distributions to account for the necessary finiteness of the moments of the number of fatalities in epidemics and in the size of other hazards.
In Sec. 3 we calculate the moments arising in those alternative approaches
and realize of the anomalously high value obtained in Ref. \cite{Cirillo_Taleb}.
We demonstrate that this is due to an unphysical divergence of the probability density of the number of fatalities at the highest-possible value.
In Sec. 4 we argue that the natural way to deal with these problems is in the context of statistical physics, using finite-size scaling,
but also show that all the proposed alternative distributions
cannot be distinguished when applied to real epidemic data, as the tail, 
which distinguishes the different distributions, is not sampled
with the number of available empirical data. 
This paper supersedes a previous unpublished short note \cite{Corral_comment_CT2}.

\section{Dual random variable, alternative transformations, and truncated power-law distribution}

\subsection{Original approach}

With the goal of solving the supposed problem of the lack of moments
of the distribution
of the number of fatalities in epidemics when this is power-law tailed, 
the authors of Ref. \cite{Cirillo_Taleb} 
introduce a map from the number of fatalities $x$ to a new random variable $z$,
called the ``dual variable'',
%that the number of fatalities $x$ is bounded 
%by the total world population $h$ (an assumption that is quite natural),
%and therefore $x$ cannot be fat-tailed.
%They further assume that
%that
%fat tails have to be studied not in $x$ (as we consider in this paper) 
%but in a transformed variable $z$ (called the dual variable),
given by the monotonic transformation
\begin{equation}
%z=\ell  +\Delta \ln\left(\frac {h-\ell }{h-x}\right),
z=\ell  - \Delta \ln\left(1-\frac {x-\ell }{h-\ell }\right),
\label{dual}
\end{equation}
where $\ell $ is the lowest value of the variable (1000 for the data of Ref. \cite{Cirillo_Taleb}),
$h$ is the world population (at the time of the epidemic),
and $\Delta=h-\ell$ or $\Delta\simeq h $ 
with no big difference in the final results as $h\gg \ell $.
The point of Ref. \cite{Cirillo_Taleb} is that the transformation from $x$ to $z$ is 
innocuous for $x\ll h$, yielding $z\simeq x$,
but ensures that $x=h$ transforms into $z=\infty$
(in practice, $z\simeq x$ even for the largest events on record).

Then, the authors \cite{Cirillo_Taleb} claim that the fat tail needs to be studied for the new variable $z$
(which is, in theory, unbounded) instead of for the measured variable $x$,
which turns out to be not fat-tailed (as desired). 
So, one fits a fat tail for $z$ and transforms back to get the distribution of $x$,
for which all moments can be easily computed, turning out to be finite,
due to the obvious fact that $\mbox{Prob}[x\le h]=1$ (as enforced by the map).
Note that,
as $z\simeq x$ for the empirical data, 
fitting a fat tail to $z$ or to $x$ leads to the same value for the exponent $\alpha$.  

The moments of $x$ calculated in this way are called the ``shadow moments'' \cite{Cirillo_Taleb},
in particular, the first moment is obtained as
%Note that the map given by Eq. (\ref{dual}) has been applied to other systems, apart from epidemics.
%These authors (ct) obtain for the first moment
%$$
%\langle x \rangle_\text{ct} = 
%(h-\mu) \left(\frac \sigma {h\xi}\right)^{1/\xi}
%\exp\left(\frac \sigma {h\xi}\right)
%\Gamma\left(1-\frac 1 \xi, \frac \sigma {h\xi}\right) + \mu
%$$
$$
\langle x \rangle_\text{ct} = 
(h-\mu) \left(\frac {\sigma\alpha} {h}\right)^{\alpha}
\exp\left(\frac {\sigma\alpha} {h}\right)
\Gamma\left(1-\alpha, \frac {\sigma\alpha} {h}\right) + \mu
$$
where $\sigma$ and $\mu$ are parameters related to the fit
when the fat tail is represented by a Pareto distribution
(the paradigmatic distribution for power-law tails in extreme-value theory \cite{Coles},
yielding a power law asymptotically \cite{Corral_Minjares})
and $\Gamma()$ is the incomplete gamma function
(we note a typo in the expression for $\langle x \rangle_\text{ct}$ in Ref. \cite{Cirillo_Taleb}
and take the expression that the same authors use in Ref. \cite{Cirillo_Taleb_QF} instead).
The subscript ct stresses that this is the result of Cirillo and Taleb \cite{Cirillo_Taleb,Cirillo_Taleb_QF}.

%EXPLICAR EL PARETO FIT!!!

%PENDIENTE!! 2:
%DECIR QUE CONSIDERAREMOS $h$ fija, el resultado para $h$ variable sera un mezcla!!!
 
%PONER AQUI LOS SHADOW MOMENTS!!!

\subsection{Other alternatives to the dual variable}

It is obvious that the dual transformation above (\ref{dual}), 
together with the fat-tailedness of $z$,
imposes an ad hoc %(ah) 
form for the tail of the number of fatalities $x$. 
But there is no justification for such an assumption [Eq. (\ref{dual})], 
and one could equally assume, for instance,
% 	QUE DOMINIO DE ATRACCION ES? DE LA GEV???
\begin{equation}
%z=\sqrt[k]{\ell ^k - \Delta^k \ln\left[1- \left(\frac{x-\ell }{h-\ell }\right)^k\right]},
z=\sqrt[k]{\ell ^k - (h^k-\ell ^k) \ln\left[1- \frac{x^k-\ell ^k}{h^k-\ell ^k}\right]},
\label{dualk}
\end{equation}
for any value of $k>0$
(with $k=1$ recovering the case considered in Ref. \cite{Cirillo_Taleb}).
This would yield, for different values of $k$, 
different ad hoc forms for the tail of the distribution of the number of fatalities $x$, 
and therefore different values for the corresponding moments, evidently.
Other choices for the map between $x$ and $z$ are possible, for example,
for $x\le h$,
$$
%z_3=x+h^\gamma\left(\frac 1 {(h-x)^\gamma}-\frac 1 {\Delta^\gamma}\right),
z=\frac h \nu \left[\frac 1{\left(1-x/h\right)^\nu}-1\right],
$$
which,
with $\nu>0$,
diverges as a power law at $x=h$.

Instead of fixing a map between $x$ and $z$ and assuming a fat tail for $z$,
one could directly assume an arbitrary distribution for $x$.
%In this way, one can 
One ``extreme'' possibility is to
consider that $x$
has a power-law shape but with a sharp truncation \cite{Burroughs_Tebbens,Corral_Serra} at $x=h$.
This truncated power-law (tpl) distribution would be given by a probability density
\begin{equation}
f_\text{tpl}(x)=\frac 1 {1-(\ell/h)^\alpha} \, \frac{\alpha \ell^\alpha}{ x^{\alpha+1}}
\label{tpl}
\end{equation}
in the range $\ell \le x < h$, and zero otherwise,
with $h > \ell > 0$ (obviously) and $\alpha\ne 0$; 
from here one could easily calculate the moments of this distribution, 
which would directly correspond to the moments of $x$ under this ad hoc prescription.
We are particularly interested in the first moment, 
which is given by
\begin{equation}
\langle x \rangle_\text{tpl} = \left(\frac{\alpha \ell}{1-\alpha}\right) \frac{(h /\ell)^{1-\alpha}-1}{1-(\ell/ h)^\alpha}.
\label{mean_tpl}
\end{equation}
This of course diverges in the limit $h\rightarrow \infty$ when $\alpha<1$.
It will be also useful to write the expression for the ccdf of the truncated power law,
$$
S_\text{tpl}(x)=\frac {\ell^\alpha} {1-(\ell/h)^\alpha} 
\left(\frac 1{x^\alpha}-\frac 1 {h^\alpha}\right)
$$
for $\ell \le x \le h$.
Below we will introduce further ad hoc options for the distribution of $x$,
in order to establish comparisons between them.

% a $z$ variable with power-law distribution truncated at $z=h$ 
%and calculate the moments of this distribution.

\section{Shadow moments 
%for alternative dual variables
and divergence of the density of fatalities at its maximum value
}

\subsection{Computation of shadow moments for different alternatives}

The number of ad hoc options for the map between $x$ and $z$ is infinite
(with a different dual variable $z$ for each map), and
%all these are ad-hoc assumptions and 
each one would lead to different values of the (shadow) moments.
In order to quantify this,
we proceed by calculating these mean values under the different ad hoc assumptions
mentioned above.
For simplicity, we will 
%use Monte Carlo simulations 
sample the power-law distribution $f_\text{pl}(z)$
and we will transform $z$ to obtain the corresponding values of $x$, using
$2\times 10^7$ realizations, restricting 
the results to $x>200,000$ fatalities; 
in other words, as in Ref. \cite{Cirillo_Taleb},
we will obtain the conditional moment $\langle x \,|\, x> 200,000\rangle$.
%but we will write it as $\langle x \rangle$ for simplicity
%(the condition on $x$ will be implicit).
The threshold for the condition was obtained in Ref. \cite{Cirillo_Taleb} as the value of $x$
from which a Pareto tail fits the data
(leading to the reported value of the exponent $\alpha=0.62$).
Note that, for the different distributions we introduce,
conditioning the moments to $x> 200,000$ (or whatever)
is equivalent to take the unconditioned moment of the 
corresponding distribution truncated at $200,000$
(i.e., replacing $\ell$ by $200,000$).

\begin{table}[h]
%\begin{centering}
\begin{center}
\caption{\label{table1} 
Mean value of $x$ conditioned to $x>200,000$
when $z$ has a power-law shape with exponent $\alpha=0.62$ 
\cite{Cirillo_Taleb} for different maps between 
$x$ and $z$, 
and when $x$ is given by a truncated power-law (tpl) distribution, 
a mixture of tpl and Dirac delta,
and a truncated gamma (tg),
as specified in the first and second columns;
for the latter distribution, Eq. (\ref{simple}), it can be verified that
the mean scales as slow as $\theta^{1-\alpha}$, as expected from Eq. (\ref{moments_fss}).
The value of $h$ is considered fixed at $h=8\times 10^9$
(for historical epidemic data this value is not fixed, 
but this is not relevant for our purposes).
%The truncated lognormal fit of Ref. \cite{Corral_epidemics_pre} as well as
The empirical value for the data of Ref. \cite{Cirillo_Taleb} is also included.
%OJO, EL VALOR TALEB HA AUMENTADO DE 2.57 A 2.65!!!
%CITAR FORMULA DE POWER LAW TRUNCADA!!!
% PENDIENTE!! POR QUE CUADRA TAN POCO EL VALOR EMPIRICO CON EL DEL FIT LOGNORMAL??
}
\smallskip
% latex_power_law_notru_col4_generic_epidemic.txt                              
\begin{tabular}
{| l|l|r|}
\hline
Model & parameters & $\langle x | x > 200,000\rangle$ \\
\hline 
Map from Eq. (\ref{dual}) \cite{Cirillo_Taleb}& $--$ & $2.65 \times 10^7$\\
%\hline
Map from Eq. (\ref{dualk}) & $k=0.1$ & $ 0.39\times 10^7$ \\
id. & $k=0.25$ & $ 1.17\times 10^7$ \\
id. & $k=0.5$ & $ 2.07 \times 10^7$ \\
% & $k=$ & $ \times 10^7$ \\
\hline
trunc. pl (tpl), Eq. (\ref{tpl}) & $--$ & $ 1.83\times 10^7$ \\ % **
\hline
delta mixture, Eq. (\ref{mixture}) & $--$ & $ 2.99\times 10^7$ \\
%\hline
%trunc. lognormal \cite{Corral_epidemics_pre} & $ 10.43$, $3.6$ & $ 6.88 \times 10^7$ \\
\hline
trunc. gamma (tg), Eq. (\ref{simple}) & $\theta=h/100$ & $0.27 \times 10^7$ \\
id. & $\theta=h/10$ & $ 0.66\times 10^7$ \\
id. & $\theta=h/2$ & $ 1.25 \times 10^7$ \\
id. & $\theta=h$ & $ 1.63 \times  10^7$ \\
id. & $\theta=2h$ & $ 2.14 \times  10^7$ \\
\hline
Empirical value \cite{Cirillo_Taleb,Corral_epidemics_pre} & $--$ & $1.40 \times 10^7$ \\
\hline
\end{tabular}
%% ** la trunc pl coincide con la formula
%	alpha=expo-1
%      AA=alpha/(1/xmin**alpha-1/h**alpha)
%	print*,alpha,\ell ,h
%	print*,'mean of the trunc PL'
%	print*, AA/(1-alpha)*(h**(1-alpha)-xmin**(1-alpha))
\par
\end{center}
%\centering{}%\caption{}
\end{table}

In the first set of simulations we assume a power-law distribution for $z$ with exponent 
$\alpha=0.62$ \cite{Cirillo_Taleb} and invert the transformation 
given by Eq. (\ref{dual}) \cite{Cirillo_Taleb} 
%as well as our generalization, Eq. (\ref{dualk}),
to get the values of $x$,
and we repeat the procedure for the generalization given by Eq. (\ref{dualk}).
Table \ref{table1} shows the results obtained for the first moment in each case;
the variability in the obtained values is apparent.
In the second set of calculations we consider that $x$ follows a power law sharply truncated (tpl) at $x=h$,
Eq. (\ref{tpl}), with the same exponent as above.
Other possibilities are explained below.
The empirical value obtained from the epidemic data of Ref. \cite{Cirillo_Taleb} is also included in the table.
%(other ad-hoc distributions for $x$ will be introduced below).
%
% 
%showing the mean of $x$, conditioned to $x>200,000$ fatalities, 
%for different assumptions for the map between $x$ and $z$
%when $z$ has a power-law tail with exponent $\alpha=0.617$ 
%(the value proposed in Ref. \cite{Cirillo_Taleb}).
%
%PUT A TABLE HERE???
%
%CONSIDERAR TAMBIEN UN SHARP CUT-OFF!!!
%

\subsection{Divergence of the Cirillo-Taleb fatality density at the maximum number of fatalities}

An illuminating result from the table is that 
the value of the (conditional) mean 
obtained from the ``extreme'' case of
a truncated power law ($1.83\times 10^7$)
is smaller than the value obtained from the map
given by Eq. (\ref{dual})
\cite{Cirillo_Taleb}
($2.65\times 10^7$). 
This %seems contradictory, 
is shocking,
as the truncated power law provides a sharp truncation of the distribution, 
and the authors of 
Ref. \cite{Cirillo_Taleb} state that the map in Eq. (\ref{dual}) provides a smooth truncation.
So, the sharpest possible truncation (at $h$),
given by the tpl,
leads to a mean value smaller 
than the pretended smooth truncation 
%of Ref. \cite{Cirillo_Taleb} (ct). 
given by Eq. (\ref{dual}).

\begin{figure}[!ht]
\begin{center}
\includegraphics[width=7.5cm]{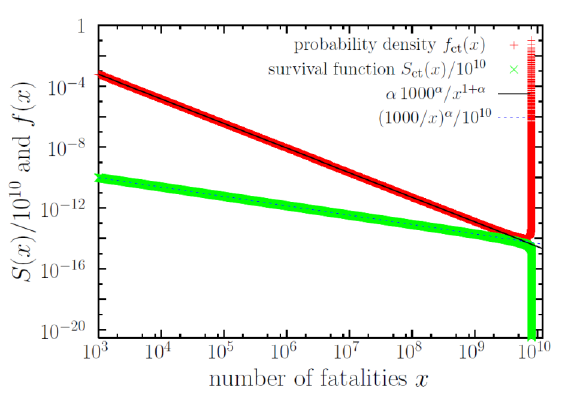}
\end{center}
\caption{
Theoretical
probability density (top curve) 
and complementary cumulative distribution function (bottom) 
of the number of epidemic fatalities $x$
obtained from a $z-$variable with a power-law shape 
%for $z\ge 1000$
to which the map 
%of Ref. \cite{Cirillo_Taleb} 
of Eq. (\ref{dual}) is applied.
%The value of the exponent, 
We use $\alpha=0.62$ \cite{Cirillo_Taleb} and $h=8\times 10^9$.
The corresponding power laws are also shown. % as straight lines.
The rapid decrease of $S_\text{ct}(x)$ at $x\rightarrow h$ translates into a divergence of $f_\text{ct}(x)$.
%Values of the survival function are shifted 10 orders of magnitude 
$S(x)$ is shifted a factor $10^{10}$ for the sake of visualization.
}
\label{fig1}
\end{figure}

To clarify this strange result,
let us calculate the %ad-hoc 
ccdf for $x$
resulting from 
the assumed power-law (pl) tail for $z$
and
the map 
%Ref. \cite{Cirillo_Taleb} 
between $z$ and $x$ 
 given by Eq. (\ref{dual}), 
$$
S_\text{ct}(x) =S_{\text{pl}}(z(x)) 
= \left( \frac \ell  {z(x)} \right)^\alpha =
 \left( \frac \ell {C-\Delta\ln(h-x)} \right)^\alpha,
$$ 
with $C$ just a constant, in concrete $C=\ell+\Delta \ln (h-\ell)$.
Note that $S_\text{ct}(h) =0$, which is necessary for normalization between $\ell$ and $h$.
We stress that for small and intermediate values of $z$ 
this variable can be directly replaced by $x$, 
but for larger values of $z$ the corresponding $x$ is stuck close to $x\simeq h$
and the ccdf $S_\text{ct}(x)$ goes to zero at $x=h$.
Figure \ref{fig1} shows the fast way in which the vanishing of $S_\text{ct}(x)$ happens. 

%PONER TAMBIEN LA PL TRUNCADA... PERO EN DENSIDAD O ACUMULADA??

%PENDIENTE!! 3: INCLUIR  MEDIA EMPIRICA!!

%PENDIENTE!!2 Explicar que la densidad tiene un minimo??

The behavior of the distribution of the resulting variable $x$ 
can be more clearly seen from its probability density.
In fact,
the relatively large change in the value of $S(x)$ in a relatively 
small range of $x$ (close to $x=h$)
can be an indication of a peak in the value of the density $f(x)$
%close to $h$
(remember that $f(x)=-dS(x)/dx$).
Indeed, one obtains
\begin{equation}
f_\text{ct}(x) = 
\frac \Delta {(h-x)}
\frac {\alpha \ell^\alpha} {[C-\Delta\ln(h-x)]^{1+\alpha}},
\label{ladivergencia}
\end{equation}
for $\ell \le x \le h$,
where the second factor tends to zero when $x\rightarrow h$, 
in the same qualitative way as $S_\text{ct}(x)$;
however, the first factor goes to infinity at $h$, 
and this is the term that dominates,
as shown in Fig. \ref{fig1} (and as one can easily calculate).
Thus, the density $f_\text{ct}(x)$ diverges as a power law at $x=h$.

This explains the high value obtained for the (conditional) mean
number of fatalities
in comparison with a truncated power law, 
as the map 
%in Ref. \cite{Cirillo_Taleb} 
of Eq. (\ref{dual})
poses an excess of mass
in the most extreme values of the distribution. 
%(which tends to a Dirac delta function).
Therefore, we conclude that the ad hoc form of the distribution
resulting from this map is clearly ``unphysical'', 
as there is no reason to assign such a high probability to ``system-size epidemics''
(pandemics killing essentially all the global population).

\subsection{No correspondence of the tail 
of the distribution arising from Eq. (\ref{dual}) with extreme-value theory}

We can compare the tail of $f_\text{ct}(x)$ with what is known in extreme-value theory \cite{Coles}.
In extreme-value theory there exists two limiting distributions, depending on the transformation one applies to the data (which depends on how one defines an extreme value).
These two types of transformations are defined by the well-known frameworks of block maxima and peaks-over-threshold;
in each case, the limiting distributions are the generalized extreme-value (GEV) distribution and the generalized Pareto distribution, respectively, 
and both are related by the tail index $\xi$, which does not change when 
one goes from one framework to the other.

As we know (decaying) power-law tails are given by $\xi > 0$; 
on the other hand,
there is a range of values of $\xi$ in which both the GEV distribution and the generalized Pareto distribution diverge at the maximum value $h$ as
$$
f_\text{evt}(x) 
\sim \left( \frac 1 {h-x}\right)^{1-1/|\xi|}
%\sim \frac 1 {(h-x)^{1-1/|\xi|}}
$$
\cite{Coles}
(with evt accounting for extreme-value theory in this case).
This happens when $\xi< -1$, for which the corresponding GEV sub-distribution is the Weibull distribution
(with the horizontal axis reversed and shifted 
with respect to the most common definition of this distribution) 
and the corresponding
sub-distribution of the generalized Pareto distribution is a special case of the beta distribution.

However, note that Eq. (\ref{dual}) leads to Eq. (\ref{ladivergencia}),
and from here,
$$
f_\text{ct}(x) \sim \frac 1 {h-x},
$$
when $x\rightarrow h$.
%at large $x$.
Therefore, this divergence at $x=h$ would correspond to the limit $\xi \rightarrow -\infty$ in the GEV and generalized Pareto distributions, which, strictly speaking, is not contemplated in standard extreme-value theory. 
Ironically, 
%the limit at $x\rightarrow h$ 
the result arising from Eq. (\ref{dual}), $\xi \rightarrow -\infty$, 
is the most opposed to a fat tail ($\xi >0$).

%PENDIENTE!! 4: MIRAR COMO VA WEIBULL EN LA COLA, 
%PERO LA DENSIDAD, O LA ACUMULADA??

%Comparison with a Weibull tail \cite{Coles} seems to indicate that
%that the distribution belongs to the so-called Weibull maximum domain of attraction, 
%with $\xi=-\infty$ (for $x$, not for $z$),
%which, ironically, is the most opposed case to a fat tail.

\subsection{Mixture of a truncated power law with a Dirac delta function}

A simple approximation is useful to clarify the approach of Ref. \cite{Cirillo_Taleb},
in particular to explain the large value of the mean of $x$.
We can approximate the map 
%of Ref. \cite{Cirillo_Taleb} 
between $x$ and $z$,
given by Eq. (\ref{dual})
by 
$x=z$ if $z < h$ and $x=h$ if $z\ge h$;
thus, all the mass of the distribution of $z$ above $z=h$ is relocated at a single point, $x=h$.
If $z$ is power-law distributed, the distribution of $x$ turns out to be a power law truncated (tpl) at $x=h$
mixed with a Dirac delta distribution at the same point $x=h$
(we label this mixed distribution by mix).

In a mathematical expression,
\begin{equation}
f_\text{mix}(x)=\left[1-\left(\frac \ell h\right)^\alpha\right] f_\text{tpl}(x) 
+ \left(\frac \ell h\right)^\alpha \delta(x-h),
\label{mixture}
\end{equation}
for $\ell \le x\le h$ and zero otherwise,
where $\delta(x-h)$ denotes a Dirac delta distribution located at $x=h$
and
the factor $(\ell/h)^\alpha$ arises because this is the weight 
$S_\text{pl}(h)$
of a power law distribution
above $z=h$
(note that the first term corresponds to $f_\text{pl}(x)$ for $x\le h$).
The first moment of this mixture of distributions will be
$$
\langle x \rangle_\text{mix}=\left[1-\left(\frac \ell h\right)^\alpha\right] \langle x \rangle_\text{tpl} 
+ \left(\frac \ell h\right)^\alpha h,
$$
and substituting the mean of truncated power law [Eq. (\ref{mean_tpl})] we get
$$
\langle x \rangle_\text{mix}=\frac{\alpha \ell}{1-\alpha} \left[\left(\frac h \ell\right)^{1-\alpha}-1\right]
+ \left(\frac \ell h\right)^\alpha h,
$$
which, with the figures used for Table \ref{table1} leads to 
$\langle x \,|\,x>200,000\rangle_\text{mix}=2.99 \times 10^7$ 
fatalities
(in agreement with the simulations).

When comparing this approximation 
with the value obtained from the map 
given by Eq. (\ref{dual})
%introduced in Ref. \cite{Cirillo_Taleb} 
(which is $2.65 \times 10^7$, see Table \ref{table1})
we see that both values are reasonably close to each other, and so, the approximation does a good job, 
and, in addition, it teaches us that, in comparison with a truncated power law, 
the map 
%of Ref. \cite{Cirillo_Taleb}
given by Eq. (\ref{dual})
subtracts a weight $(\ell/h)^\alpha$ from a tpl distribution defined between $\ell$ and $h$ and puts 
all this weight around $x=h$, see Eq. (\ref{mixture}).
This is the reason why $\langle x \rangle_\text{ct}$ is larger than 
$\langle x \rangle_\text{tpl}$ (but somewhat smaller than $\langle x \rangle_\text{mix}$).
In any case, this shows how one cannot consider the truncation 
given by Eq. (\ref{dual})
\cite{Cirillo_Taleb} as ``smooth''.

It is illustrative to write down the expression of the ccdf in this case (mix),
to compare it with those of the power law and the truncated power law. 
%Eq. (\ref{tpl}). 
The former is
$$
S_\text{mix}(x)=\left(\frac \ell x\right)^\alpha \Theta(h-x),
%\S
$$
for $x\ge \ell$,
where we have explicitly included in the expression the Heaviside step function, $\Theta()$,
in order to clarify that the Dirac delta function in the density comes from the derivative of $\Theta$.
Some of the expressions above can also be written using $\Theta()$,
for instance, Eq. (\ref{tpl}), 
$f_\text{tpl}(x) \propto \alpha \ell^\alpha \Theta(h-x)/x^{\alpha+1}$ ,
but note the fundamental difference between including such a truncation in $S(x)$
or in $f(x)$
(for $f(x)$ that truncation is sharp, 
but for $S(x)$ it is ``worse'' than sharp).

%ENTONCES ES GUMBEL O WEIBULL??
%
%FOOLED BY THE USE OF THE SURVIVAL FUNCTION!!
%

%IF THE MEAN OF THE DISTRIBUTION IS NOT WELL DEFINED DON'T USE THE MEAN!!!!

%DEFINIR EL RANGO DE LOS PARAMETROS K Y BETA y el rango de la variable x

\section{Finite-size scaling and incomplete sampling of the tail}

\subsection{Finite-size scaling}
%Let us set clearly that,
In statistical physics, the ``canonical'' approach for this kind of problems
(when one expects finite-size effects due to the finite size of the system)
is a finite-size scaling (fss) ansatz.
In this framework, one can write for the probability density
$$
%\begin{equation}
f_{\text{fss}}(x)=\frac 1 \ell  \left(\frac \ell  x \right)^{1+\alpha} G\left(\frac x {L^d}\right),
%\label{threebis}
%\end{equation}
$$
for $x\ge \ell$ (and zero otherwise),
where 
$L$ is a measure of system size,
$d$ is a (positive) scaling exponent,
and $G$ an unknown scaling function, 
whose shape has to be constant for small arguments and decaying fast 
%(for example, exponentially) 
for large arguments, see Ref. \cite{Corral_csf}.
The exact formula for $G$ can be rather complicated for very simple models, 
see for instance Ref. \cite{Corral_garcia_moloney_font}.
$G$ does not have to be confused with the slowly varying function
%$\ell(x)$ 
defining fat tails \cite{Voitalov_krioukov}; 
in fact, the scaling function destroys fat-tailedness.

Note that by system size we refer to a spatial size,
and thus, $L$ is typically a measure of the extend of the system.
The term ``finite size'' refers to the finiteness of $L$, 
and not to the random variable $x$ 
(which is also a ``size'', although in our case this size is measured as number of fatalities). 
The idea of finite-size scaling is that 
the finiteness of the size of the system $L$ 
limits the growth of the size $x$ of the phenomenon.
Although the finite-size scaling hypothesis may seem another arbitrary assumption, 
it is supported by an enormous amount
of theory and simulation results \cite{Privman,GarciaMillan}.
It has the advantage that it is largely non-parametric, 
as the scaling function $G$ is kept unknown.

Within this framework 
one cannot obtain the exact moments of $x$ 
(as the scaling function is unknown)
but one can obtain that the moments scale as
\begin{equation}
\langle x^q \rangle_\text{fss} 
\propto 
\left\{
\begin{array}{ll}
{\ell ^{q}} &\mbox{ for } q < \alpha,\\
{\ell ^{\alpha}} {L^{d(q-\alpha)}} &\mbox{ for } q > \alpha,\\
\end{array}
\right.
\label{moments_fss}
\end{equation}
%OJO, CUAL ES EL RANGO DE ALPHA???
with $\alpha >0$ for proper normalization \cite{Corral_csf}.
This equation sets clearly the divergence of the moments for $q>\alpha$
when $L\rightarrow \infty$,
for which fat-tailedness is recovered.
In the case of the mean, when $\alpha < 1$, we have that
$\langle x \rangle_\text{fss} 
\propto 
{\ell ^{\alpha}} {L^{d(1-\alpha)}}$.
Apart from the knowledge of $\alpha$,
the key ingredient to obtain the right scaling of the moments (not their exact expression) 
is the determination of the exponent $d$.

Note that one could naively assume, from wrong dimensional analysis, 
that $h\propto L^d$,
but there is no reason for such an assumption
(we reinterpret $h$ not as the world population but as the population of the region
of size $L$ under consideration).
%this is not the case in many well studied models \cite{Corral_garcia_moloney_font}.
The point is that $d$ reflects the way in which the process
(the epidemic in the case we are concerned) 
``fills'' the underlying space 
%or population
(assuming that the number of fatalities is proportional to the number of cases).

For instance, let us consider a process of percolation in the Bethe lattice,
%(also called Cayley tree), 
equivalent to the Galton-Watson branching process \cite{Christensen_Moloney,Corral_garcia_moloney_font}.
For different system sizes $L$, 
with $L$ corresponding to the depth of the considered tree,
it turns out that, at the critical point, $d=2$, i.e., $x$ scales as $L^2$;
however, the total population $h$ (the underlying tree) grows exponentially with $L$,
which means that the process, or epidemic, is very far from ``filling'' the whole population
(the possibility that epidemics tend to be close to their critical point is discussed at 
Ref. \cite{Corral_epidemics_pre}).
A less dramatic example happens in two-dimensional site percolation
\cite{Christensen_Moloney}, 
for which $d=1.896$, which is significantly smaller than the spatial dimension 2.

In reality, the situation is somewhat more complex.
Although we can assume that the number of fatalities scales linearly with
(i.e., is proportional to) the number of cases,
the number of cases does not necessarily scale linearly with the population
(as the examples above illustrate).
But, further, the population does not scale trivially (as $L^2$) with space,
due to the fractal structure of human population \cite{Corral_Muro}.
In other words, 
we expect a sublinear growth of $x$ with total population,
and a sublinear growth of the total population with $L^2$,
and therefore we conclude that the exponent $d$ has to be smaller than $2$.
In any case, 
%In other words, the finite world population affects the distribution of $x$ in a non-linear, non-trivial way, leading to a cutoff $L^d \ll h$, and thus, 
$h$ turns out to be a rather bad upper bound for $x$
(contrary to what is assumed in Ref. \cite{Cirillo_Taleb}).

%In the case of epidemics, the value of the exponent $d$ is unknown.

A simplified, concrete option \cite{Serra_Corral} can be
to consider that $G$ is a decreasing exponential, so,
\begin{equation}
f_{\text{tg}}(x) \propto 
\frac 1 \ell  \left(\frac \ell  x \right)^{1+\alpha} e^{-x/\theta},
\label{simple}
\end{equation}
for $x\ge \ell$ (and zero otherwise),
with $\theta$ a scale parameter verifying $\theta\propto L^d$.
This corresponds to a truncated gamma (tg) distribution
(note that, in contrast to the truncated power law,
where the truncation refers to $x\le h$,
the truncated gamma refers to truncation from below, $x\ge \ell$).

We have argued that we expect $\theta=L^d \ll h$,
however, for the sake of the argument, 
we are going to consider now several values of $\theta$ close to
(and even larger than) $h$.
Results from simulations 
%%of a particular case with $d=1$ amd 
%for
%several values of $\theta$ 
%%(leaving out the value of $d$)
appear in Table \ref{table1}.
Even for large values of $\theta$ (as $\theta=2h$)
the resulting $\langle x \rangle_\text{tg}$
is smaller than $\langle x \rangle_\text{ct}$.
This is remarkable, as the tg distribution with $\theta=2h$
is unrealistic, as it gives a non-negligible probability that $x$ is above twice
the world population, but despite this lack of reality it yields a first moment smaller
than the one for the ct case (the one arising from Eq. (\ref{dual})).
Notice that
an important advantage of the truncated gamma with respect the truncated power law
and the distributions arising for the different maps, as Eq. (\ref{dual}),
is that the later %(in its most natural implementation)
implicitly assume $h\propto L^d$,
but we have already mentioned that this is a bad upper bound.

In practice, with the available epidemic data 
(which only includes total number of fatalities)
one cannot perform a finite-size scaling study,
as no spatial information is available.
Ideally,
one would need the spatial coordinates of each fatality,
and the empirical distribution $f(x)$ should be computed for different subsystems of size $L$
%(notice that this spatial data is not even available for the ). 
In that way, one could see how the number of fatalities changes with the area under observation, similarly as was done in Ref. \cite{Corral_Muro}
for a different problem.
One should assume that the critical properties of different epidemics should be equivalent, 
i.e., universal.
In any case, although the value of $d$ cannot be estimated from data, 
our approach shows (Table \ref{table1} for different values of $\theta$) 
that, no matter the value of $d$,
the expected number of fatalities with the finite-size scaling assumption should be smaller than the one obtained from the ad hoc assumption given by Eq. (\ref{dual}) \cite{Cirillo_Taleb}.

\subsection{Undersampling of the tail}

\begin{figure}[!ht]
\begin{center}
\includegraphics[width=7.5cm]{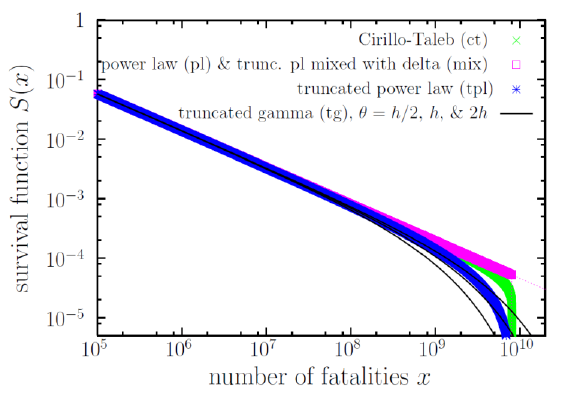}\\
\includegraphics[width=7.5cm]{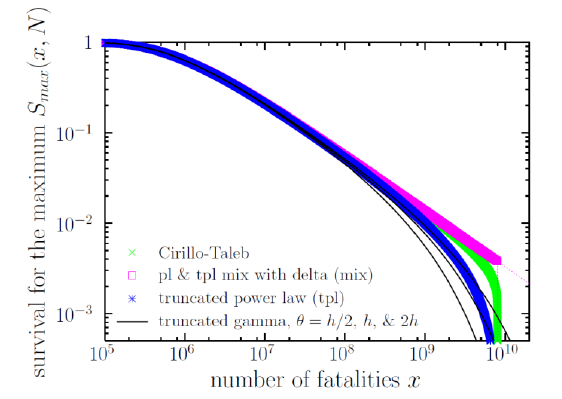}\\
%C:\Users\Alvaro Corral\OneDrive - UAB\projects\pandemic_prog_data\prog_new
%gnuplot> load'all_distributions.grf       
\end{center}
\caption{
(a) Complementary cumulative distribution functions (from top to bottom) for the 
power law, 
the truncated power law mixed with a Dirac delta at $x=h$,
the Cirillo-Taleb (ct) \cite{Cirillo_Taleb} case,
and the truncated power law, 
with $\alpha=0.62$ \cite{Cirillo_Taleb} and $h=8\times 10^9$.
The two first distributions are represented by the same curve
up to $x=h$,
but the power law should be extended to infinity (with the same slope)
whereas the mixed case should go sharply to zero at $x=h$.
Truncated gamma distributions, for different $\theta$, are also displayed
(thinner lines).
(b) Same for the probability that at least one value is above $x$ when $N=72$.}
%as in Ref. \cite{Cirillo_Taleb}.}
\label{cumulative_dists}
\end{figure}

It is clear that the different distributions proposed here
(together with the distribution implicit in Eq. (\ref{dual}) \cite{Cirillo_Taleb})
are only different for the largest values of the variable;
however, these largest values are not sampled when the number of data $N$ is not large enough.
Figure \ref{cumulative_dists}(a) compares the ccdf
when $\alpha=0.62$
%for the ct case \cite{Cirillo_Taleb}, the power-law case (pl), the tpl mixed with a delta (mix),
%and the truncated power law (tpl),
for the power law, 
the truncated power law mixed with a Dirac delta, 
%at $x=h$,
the Cirillo-Taleb (ct) \cite{Cirillo_Taleb} case (Eq. (\ref{dual})),
and the truncated power law,
showing that the four distributions are nearly the same up to $x\simeq 10^9$, 
for which $S(x)\simeq 2 \times 10^{-4}$.
In fact, it is the tpl distribution that departs from the rest at this point, 
and the latter are nearly indistinguishable up to $x\simeq 5 \times 10^9$,
corresponding to $S(x) < 10^{-4}$.
The plot also includes the truncated gamma distribution with different values of the scale parameter $\theta$; in this case the distributions depart from the power law at smaller values of $x$ (at $S(x)\simeq 10^{-3}$ for $\theta=h/2$) and the decay is much slower.

We can quantify this further by calculating the probability that the tail of the distribution is sampled
when the number of available data is $N$.
The probability of ``seeing'' the tail, starting at $y$ (by convention) 
is the probability that at least one value of $x$ is above $y$, and this is given (assuming independence) by
$$
S_\text{max}(y,N)=1-[1-S(y)]^N.
$$
Note that this is the same as the probability that the maximum of $N$ values is above $y$
and thus we refer to it as $S_\text{max}(y,N)$
(empirically, $S_\text{max}(y,N)$ is given by a Heaviside step function with its jump at the maximum recorded value of $x$).

Figure \ref{cumulative_dists}(b) plots $S_\text{max}(y,N)$ for $N=72$ 
(the number of data used in Ref. \cite{Cirillo_Taleb}), 
and it can be seen how for the pl, mix, ct, and the tpl cases, the probability of getting a value of $x$ above $10^9$
(where the distributions start to differ) is around 0.01 (somewhat smaller for the tpl case).
This means that we could see that (single) value in 1 out of 100 realizations, 
but, of course, we have only one empirical realization in the case of the statistics of epidemics.
Thus, the purpose of distinguishing between the different distributions is futile 
when $N$ is as low as $72$. 
%(which corresponds to the empirical situation in epidemics).
Including in the comparison the truncated gamma case
one can see that the departure from the rest of the models takes place
for somewhat smaller values of $x$, giving a value of $S_\text{max}$ about 0.05 for $\theta=h/2$,
which makes the distinction of these different distributions still impossible in practice.

%In order to compare with the statistics of epidemics, 
%Ref. \cite{Corral_epidemics_pre} suggested that,
%even if the preferred fit was lognormal,
% an ``acceptable'' power-law fit
%is given by a tail starting at $\ell=33,000$ fatalities, 
%with $\alpha=0.34$.
%Figure XXX compares this power-law tail with the equivalent distributions
%in the finite-size scaling case, taking the truncated-gamma example.
%PENDIENTE!!

\section{Conclusions}

We have addressed the apparent problem of the lack of moments 
when using power-law tailed distributions to model extreme events in epidemics
(and other hazards), 
and in particular, the divergence of the mean when the exponent of the complementary cumulative distribution function 
is $\alpha < 1$.
The authors of
Ref. \cite{Cirillo_Taleb}
approached the problem in the context of the number of fatalities of epidemics
(and in the context of wars and finance in Refs. \cite{Cirillo_Taleb_wars,Cirillo_Taleb_QF}),
introducing a transformation between the original variable $x$ (the number of fatalities) and a dual variable $z$, given by Eq. (\ref{dual}).
The transformation ensured that even if $z$ is unbounded (as corresponds to a power-law tailed variable), $x$ does not surpass a maximum value $h$ (the world population in the case of epidemics).
This dual framework allowed to calculate the moments of $x$,
which obviously turn out to be all finite.

Here
we have shown that the transformation given by Eq. (\ref{dual})  \cite{Cirillo_Taleb}
is ad hoc, having no physical justification.
Of course, other transformations are equally possible, as illustrated for instace by Eq. (\ref{dualk}), 
and these alternative transformations lead to values of the moments
that are different from the ones obtained in Ref. \cite{Cirillo_Taleb}.
In fact, introducing an ad hoc dual transformation is equivalent to introduce 
an ad hoc form for the function modeling the distribution of fatalities.
We have considered illustrative to consider the truncated power law,
the truncated power law mixed with a Dirac delta function,
and the truncated gamma.

But
we further argue that the problem has to be put in the context of statistical physics, 
where the proper way to deal with finite-size effects is by means of finite-size scaling
(due to the obvious fact that
the finite %world population 
size $L$ of the planet
has to introduce finite-size effects in a hypothetical power-law tailed distribution of the number of epidemic fatalities).
However, due to the scarcity of epidemic data
and the lack of spatial information,
one cannot implement an empirical finite-size scaling approach.
An alternative option, 
thanks to the universality paradigm \cite{Stanley_rmp},
 would be to use computer simulations, 
%with different values of the total population $h$, 
with a model in the same universality class than real epidemics.
Needless to say, this universality class is unknown nowadays, 
and so, the exponent $d$ and therefore the scaling of the moments of the distribution will remain unknown as well.

%ARRIBA MAS ROLLO???

Comparing the moments arising from the different alternative distributions 
we arrive to the 
fact that the mean value obtained in Ref. \cite{Cirillo_Taleb} is surprisingly high
(in comparison with the other distributions).
This is counter-intuitive, as the authors of Ref. \cite{Cirillo_Taleb}
claim that their truncation is smooth, 
and we use a very sharp truncation in the case of the truncated power law,
which however leads to a smaller mean value.

We find that the reason of this disagreement is that the truncation provided by 
the dual transformation is not smooth at all, 
but enormously sharp, as it accumulates all the mass coming from $z>h$
around the point $x\simeq h$,
yielding a diverging peak of the density at $x=h$.
This tail is so sharp that it is not included in the well-know domains of attraction of the 
generalized extreme-value distribution \cite{Coles}.
%(corresponding to $\xi \rightarrow -\infty$).
The fact that the distribution arising from 
Eq. (\ref{dual})
%Ref. \cite{Cirillo_Taleb}
seems to be smoothly truncated when looking at the ccdf $S_\text{ct}(x)$
but it is so anomalous when looking at its probability density $f_\text{ct}(x)$
 illustrates how the characterization of a distribution by its ccdf is misleading, 
and the probability density provides a more intuitive picture.

From our analysis we conclude that the fundamental 
problem in the statistics of epidemics
and other natural hazards governed by power-law tails is not the divergence of moments
but the fact that the real tail of the empirical distribution is not sampled in practice, 
as the number of empirical data $N$ is too low.
The lesson learned from here is that, as moments cannot be constrained, 
it is pointless to try to calculate them and one should rely on other metrics 
to assess risk.
Otherwise, what one gets is not any characterization of the empirical data but a reflection of the properties of the ad hoc model chosen.

\section*{Acknowledgments}

%I acknowledge my colleague Isabel Serra for discussions
%and
%support from projects
%FIS2015-71851-P and
%PGC-FIS2018-099629-B-I00
%Red de Excelencia MAT2015-69777-REDT, 
%and 
%Mar\'{\i}a de Maeztu Program 
%%for Units of Excellence in R\&D 
%MDM-2014-0445,
%from Spanish MINECO and MICINN.

The Centre de Recerca Matem\`atica is supported by the CERCA Programme of the Generalitat de Catalunya as well as by the Spanish State Research Agency (AEI) through the Severo Ochoa and Mar\'{\i}a de Maeztu Program for Centers and Units of Excellence in R\&D (CEX2020-001084-M).
The research of the author is supported by the projects PGC-FIS2018-099629-B-I00
and PID2021-125979OB-I00, AEI.

%\bibliographystyle{unsrt}   % por orden de cita
%\bibliography{C:/Users/acorr/Dropbox/p1_lemmas/biblio} % portatil
%%%\bibliography{C:/Users/acorral/Dropbox/p1_lemmas/biblio}

\begin{thebibliography}{10}

\bibitem{Bernoulli_smallpox}
D. Bernoulli, S. Blower.
\newblock An attempt at a new analysis of the mortality caused by smallpox and
  of the advantages of inoculation to prevent it.
\newblock {\em Reviews in Medical Virology}, 14(5):275--288, 2004.

\bibitem{Dietz_Bernoulli}
K.~Dietz and J.~A.~P. Heesterbeek.
\newblock {Daniel Bernoulli}’s epidemiological model revisited.
\newblock {\em Math. Biosci.}, 180(1):1--21, 2002.

\bibitem{Pastor_Vespignani}
R.~Pastor-Satorras and A.~Vespignani.
\newblock Epidemic spreading in scale-free networks.
\newblock {\em Phys. Rev. Lett.}, 86:3200--3203, 2001.

\bibitem{Kleinberg_book}
D.~Easley and J.~Kleinberg.
\newblock {\em Networks, Crowds, and Markets}.
\newblock Cambridge Univ. Press, 2010.

\bibitem{Pastor_rmp}
R.~Pastor-Satorras, C.~Castellano, P.~{Van Mieghem}, and A.~Vespignani.
\newblock Epidemic processes in complex networks.
\newblock {\em Rev. Mod. Phys.}, 87:925--979, 2015.

\bibitem{Allen}
L.~J.~S. Allen.
\newblock A primer on stochastic epidemic models: Formulation, numerical
  simulation, and analysis.
\newblock {\em Infectious Disease Modelling}, 2(2):128--142, 2017.

\bibitem{Miller_pgf}
J.~C. Miller.
\newblock A primer on the use of probability generating functions in infectious
  disease modeling.
\newblock {\em Infectious Disease Modelling}, 3:192--248, 2018.

\bibitem{Kucharski_book}
A.~Kucharski.
\newblock {\em The rules of contagion}.
\newblock Basic Books, London, 2020.

\bibitem{Hill_epidemics}
A.~L. Hill.
\newblock The math behind epidemics.
\newblock {\em Phys. Today}, 73(11):28--34, 2020.

\bibitem{Arenas_covid}
A.~Arenas, W.~Cota, J.~G\'omez-Garde\~nes, S.~G\'omez, C.~Granell, J.~T.
  Matamalas, D.~Soriano-Pa\~nos, and B.~Steinegger.
\newblock Modeling the spatiotemporal epidemic spreading of {COVID-19} and the
  impact of mobility and social distancing interventions.
\newblock {\em Phys. Rev. X}, 10:041055, 2020.

\bibitem{Castro_Ares_PNAS20}
M.~Castro, S.~Ares, J.~A. Cuesta, and S.~Manrubia.
\newblock The turning point and end of an expanding epidemic cannot be
  precisely forecast.
\newblock {\em Proc. Natl. Acad. Sci. USA}, 117(42):26190--26196, 2020.

\bibitem{Falco_Corral}
C.~Falcó and A.~Corral.
\newblock Finite-time scaling for epidemic processes with power-law
  superspreading events.
\newblock {\em Phys. Rev. E}, 105:064122, 2022.

\bibitem{Cirillo_Taleb}
P.~Cirillo and N.~N. Taleb.
\newblock Tail risk of contagious diseases.
\newblock {\em Nature Phys.}, 16:606--613, 2020.

\bibitem{Wikipedia_epidemics}
Wikipedia.
\newblock List of epidemics.
\newblock {\em https://en.wikipedia.org/wiki/List$\_$of$\_$epidemics}.

\bibitem{footnote1_comment2_CT}
To avoid confusion, let us mention that in survival analysis $S(x)$ accounts
  for the lifetime $x$ of an individual, but this is not the case here;
  nevertheless, the idea is essentially the same if one counts the ``lifetime''
  of an epidemic not using time but number of fatalities instead.

\bibitem{Voitalov_krioukov}
I.~{Voitalov}, P.~{van der Hoorn}, R.~{van der Hofstad}, and D.~{Krioukov}.
\newblock Scale-free networks well done.
\newblock {\em Phys. Rev. Research}, 1:033034, 2019.

\bibitem{Corral_epidemics_pre}
A.~Corral.
\newblock Tail of the distribution of fatalities in epidemics.
\newblock {\em Phys. Rev. E}, 103:022315, 2021.

\bibitem{Bouchaud_Georges}
J.-P. Bouchaud and A.~Georges.
\newblock Anomalous diffusion in disordered media: statistical mechanisms,
  models and physical applications.
\newblock {\em Phys. Rep.}, 195:127--293, 1990.

\bibitem{Mantegna_Stanley}
R.~N. Mantegna and H.~E. Stanley.
\newblock {\em An Introduction to Econophysics}.
\newblock Cambridge Univ. Press, Cambridge, UK, 2000.

\bibitem{Cirillo_Taleb_wars}
P.~Cirillo and N.~N. Taleb.
\newblock On the statistical properties and tail risk of violent conflicts.
\newblock {\em Physica A: Statistical Mechanics and its Applications},
  452:29--45, 2016.

\bibitem{Cirillo_Taleb_QF}
P.~Cirillo and N.~N. Taleb.
\newblock Expected shortfall estimation for apparently infinite-mean models of
  operational risk.
\newblock {\em Quantitative Finance}, 16(10):1485--1494, 2016.

\bibitem{Clauset}
A.~Clauset, C.~R. Shalizi, and M.~E.~J. Newman.
\newblock Power-law distributions in empirical data.
\newblock {\em SIAM Rev.}, 51:661--703, 2009.

\bibitem{Corral_Deluca}
A.~Deluca and A.~Corral.
\newblock Fitting and goodness-of-fit test of non-truncated and truncated
  power-law distributions.
\newblock {\em Acta Geophys.}, 61:1351--1394, 2013.

\bibitem{Corral_Gonzalez}
A.~Corral and A.~Gonz\'alez.
\newblock Power law distributions in geoscience revisited.
\newblock {\em Earth Space Sci.}, 6(5):673--697, 2019.

\bibitem{Corral_comment_CT2}
A.~Corral.
\newblock Finite-size scaling versus dual random variables and shadow moments
  in the size distribution of epidemics.
\newblock {\em arXiv}, 2011.04316, 2020.

\bibitem{Coles}
S.~Coles.
\newblock {\em An Introduction to Statistical Modeling of Extreme Values}.
\newblock Springer, London, 2001.

\bibitem{Corral_Minjares}
A. Corral, M. Minjares, and M. Barreiro.
\newblock Increased extinction probability of the Madden-Julian oscillation after about 27 days.
\newblock {\em Phys. Rev. E}, 108:054214, 2023.

\bibitem{Burroughs_Tebbens}
S.~M. Burroughs and S.~F. Tebbens.
\newblock Upper-truncated power laws in natural systems.
\newblock {\em Pure Appl. Geophys.}, 158:741--757, 2001.

\bibitem{Corral_Serra}
A.~Corral and I.~Serra.
\newblock Time window to constrain the corner value of the global
  seismic-moment distribution.
\newblock {\em PLoS ONE}, 14(8):e0220237, 2019.

\bibitem{Corral_csf}
A.~Corral.
\newblock Scaling in the timing of extreme events.
\newblock {\em Chaos. Solit. Fract.}, 74:99--112, 2015.

\bibitem{Corral_garcia_moloney_font}
A.~Corral, R.~Garcia-Millan, N.~R. Moloney, and F.~Font-Clos.
\newblock Phase transition, scaling of moments, and order-parameter
  distributions in {Brownian} particles and branching processes with
  finite-size effects.
\newblock {\em Phys. Rev. E}, 97:062156, 2018.

\bibitem{Privman}
V.~Privman.
\newblock Finite-size scaling theory.
\newblock In V.~Privman, editor, {\em Finite Size Scaling and Numerical
  Simulation of Statistical Systems}, pages 1--98. World Scientific, Singapore,
  1990.

\bibitem{GarciaMillan}
R.~Garcia-Millan, F.~Font-Clos, and A.~Corral.
\newblock Finite-size scaling of survival probability in branching processes.
\newblock {\em Phys. Rev. E}, 91:042122, 2015.

\bibitem{Christensen_Moloney}
K. Christensen and N. R. Moloney.
\newblock {\em Complexity and Criticality}.
\newblock Imperial College Press, London, 2005.

\bibitem{Corral_Muro}
A. Corral and M. Garc\'{\i}a del Muro.
\newblock Finite-size scaling of human-population distributions over fixed-size cells and its relation to fractal spatial structure.
\newblock {\em Phys. Rev. E}, 106:054310, 2022.

\bibitem{Serra_Corral}
I.~Serra and A.~Corral.
\newblock Deviation from power law of the global seismic moment distribution.
\newblock {\em Sci. Rep.}, 7:40045, 2017.

\bibitem{Stanley_rmp}
H.~E. Stanley.
\newblock Scaling, universality, and renormalization: {Three} pillars of modern
  critical phenomena.
\newblock {\em Rev. Mod. Phys.}, 71:S358--S366, 1999.

\end{thebibliography}

\end{document}